\documentclass[12pt,preprint]{aastex}

\slugcomment{}

\begin{document}

\title{A flux emergence model for solar eruptions}

\author{V. Archontis\altaffilmark{1}
and A.W. Hood\altaffilmark{1}}
\altaffiltext{1}{School of Mathematics and Statistics, University of St
  Andrews, North Haugh, St Andrews,  Fife KY16 9SS, UK}

\begin{abstract}
We have simulated the 3D emergence and interaction of two twisted flux tubes, which
rise from the interior into the outer atmosphere of the Sun.
We present evidence for the multiple formation and eruption of flux ropes inside the emerging flux systems and hot arcade-like 
structures in between them. Their formation is due to internal reconnection, occurring between oppositely 
directed, highly stretched and sheared fieldlines at photospheric heights.
Most of the eruptions escape into the corona, 
but some are confined and fade away without leaving the low atmosphere. As these flux ropes erupt, new 
reconnected fieldlines accumulate around the main axis of the initial magnetic flux systems.
We also show the complex 3D fieldline geometry and the structure of the multiple current sheets, which 
form as a result of the reconnection between the emerging flux systems.
\end{abstract}


\section{Introduction}

Partial eruption of a flux rope can be produced by favourable, \textit{imposed} shearing motions 
at the photosphere. This shearing can occur as sunspots with opposite 
polarity move apart along the neutral line of an active region. Then, as the rising 
magnetic system expands, the sheared magnetic fieldlines are stretched and undergo 
internal reconnection, at low heights, which leads to partial eruption of magnetic flux 
\citep{Manchester+ea04}. This is consistent with the `tether-cutting'
mechanism (\citealt{Sturrock89, Moore01}), in which, effective reconnection of 
magnetic fieldlines at the core of  a magnetic bipole releases the energy stored in the 
sheared field, resulting in an eruption.
An important factor in determining whether an emerging flux rope will erupt 
is the amount of twist of the fieldlines around the central axis of the tube (\citealt{Murray06}). 
If the winding of the fieldlines is high enough, a kink 
instability occurs that eventually leads to either a confined or an ejective eruption 
(\citealt{Torok05, Gibson+ea06}).

Observations (\citealt{Sterling04, Sterling05, Jiang07}) have shown active-region filament eruptions 
when new flux emerges near the location of a preexisting emerging flux region. It is likely that 
the new emergence is important in the eruption onset. Several numerical 
models, requiring multipolar magnetic configurations for eruptions, have 
reproduced the above and similar-like observations. In the breakout model 
(\citealt{Antiochos98}), a sheared bipole rises into an overlying bipole and \lq external\rq\
reconnection occurs at a magnetic null point at their interface. When the reconnection becomes fast 
enough the 
rising bipole breaks through the ambient field and erupts.

Another possibility, shown in magnetofrictional simulations, is that flux ropes are formed and 
erupt due to slow diffusion of 
decaying flux from active regions, which interact through converging motions on the photosphere 
(\citealt{Duncan06, Adrian07}).
Also, \citet{Chen00} simulated the interaction between a flux system with a detached flux 
rope and an emerging flux system that appears to one side of the primary field. Reconnection between 
the flux systems leads to a vertical current sheet below the detached flux rope, which eventually 
experiences a fast ejection.

In this Letter, we consider the 3D extension of the model by \citet{Archontis07}
(hereafter called AHB model)
in which one (leading) flux tube rises through a stratified atmosphere to create a coronal 
field for the second (following) tube to rise into. The 3D model simulates: 1) the intricate 
reconnection of the two flux systems 2) the multiple eruptions of flux ropes, which are initiated
by the new emergence of flux and 3) the formation of current sheets and hot arcade structures.

\section{Model}

The initial stratification of the stratified atmosphere and the properties of the magnetic flux 
tubes are similar to the AHB model. 
The simulations use a 3D version of the Lare shock-capturing code 
(see  \citet{Arber01}). The time-dependent, resistive and compressible MHD 
equations are solved by a Lagrangian remap scheme, including uniform gravity. 
The energy equation is adiabatic and includes viscous and Joule dissipation terms. 
The resistivity is enhanced at sites where the current density exceeds a certain
value in a smilar manner to the AHB model.
At the beginning of the experiment two flux tubes are located $2.9$Mm and $2.2$Mm below the photosphere, 
with axes parallel to the $y$ direction. The fieldlines are uniformly twisted around the main axis of each 
tube. The initial twist is such that the tubes are stable to the kink instability. Both tubes are left-handed and can reconnect when they come into contact. The rising motion of the tubes is triggered by a density excess 
along their central axes. The density deficit is reduced away from $y=0$ following a Gaussian 
profile, similar to the experiment by \citet{Archontis05}. The length scale of the buoyant part of the 
tubes is defined by the parameter $\lambda$, where $\lambda=20$ in our experiments.
This deficit makes the tubes less dense and more buoyant in the middle part of their 
axes. As they rise through the solar interior, they form $\Omega$-shaped loops due to 
magnetic buoyancy.

All variables are made dimensionless against their photospheric values, namely: 
pressure, $P=1.4\,10^5$ erg cm$^{-3}$;
density, ${\rho}=3\,10^{-7}$ g cm$^{-3}$; temperature, $T=5.6\,10^3$ K;
scale height, $H=170$ km;
time, $t=25$ sec; velocity, $V=6.8$ km sec$^{-1}$ and magnetic field, $B=1.3\,10^3$ Gauss.
The dimensionless coordinates are $(-100 \le x \le 100)$, $(-100 \le y \le 100)$ and $(-10 \le z \le 140)$.
The stratified atmosphere includes the top of the solar interior from $z=-10$ to $22$,
two isothermal layers ($z=22$ to $32$ for the photosphere and $z=42$ to $140$ for the
corona) and a {\it transition region} with a steep temperature gradient joining them.
The size of the grid used in the numerical experiments is
$(256,256,320)$ points in the $(x,y,z)$ directions,
with $x$ being the transverse coordinate and $z$ the vertical coordinate.

\section{Results and Discussion}
The initial evolution of the system is similar to the evolution in the 
AHB model.
When the leading tube reaches the photosphere a bipolar region is formed with a north-south 
orientation due to the initial strong twist of the fieldlines. Eventually, the bipolar region 
moves towards an east-west direction as more internal magnetic layers rise to the photosphere and 
the inclination of the anchored flanks of the tube becomes more vertical. 
After $t=35$, the leading tube rises into the non-magnetized 
corona where it expands and \textit{creates an ambient field} for the following tube 
to rise into. The direction of the fieldlines and the field strength vary away 
from the center of the expanding volume of the leading system and, thus, the newly 
formed ambient field is more complex than in the previous simulations
of the interaction between an emerging field and a uniform coronal field.

\begin{figure*}[htb]
\includegraphics[scale=0.18]{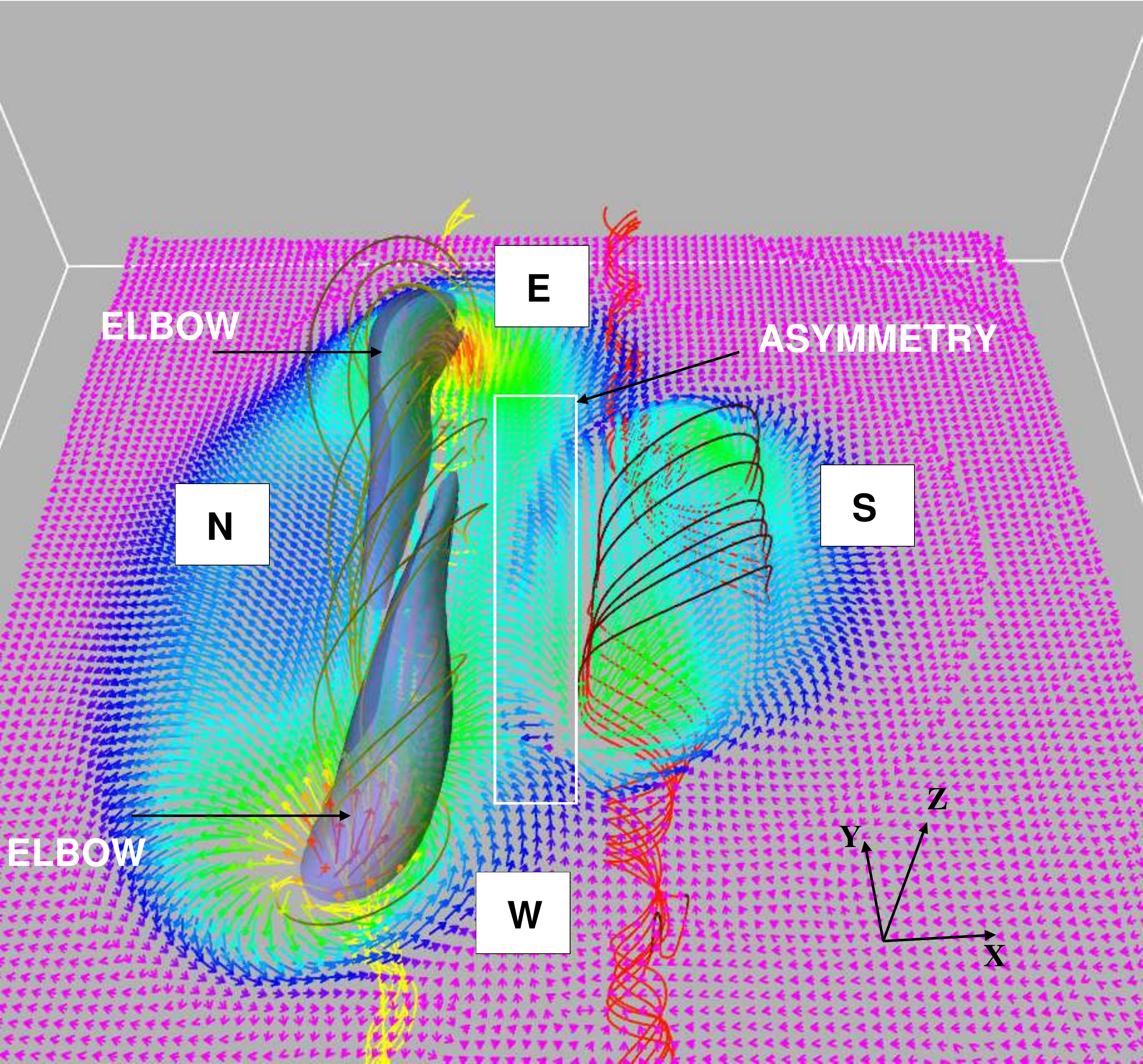}
\includegraphics[scale=0.18]{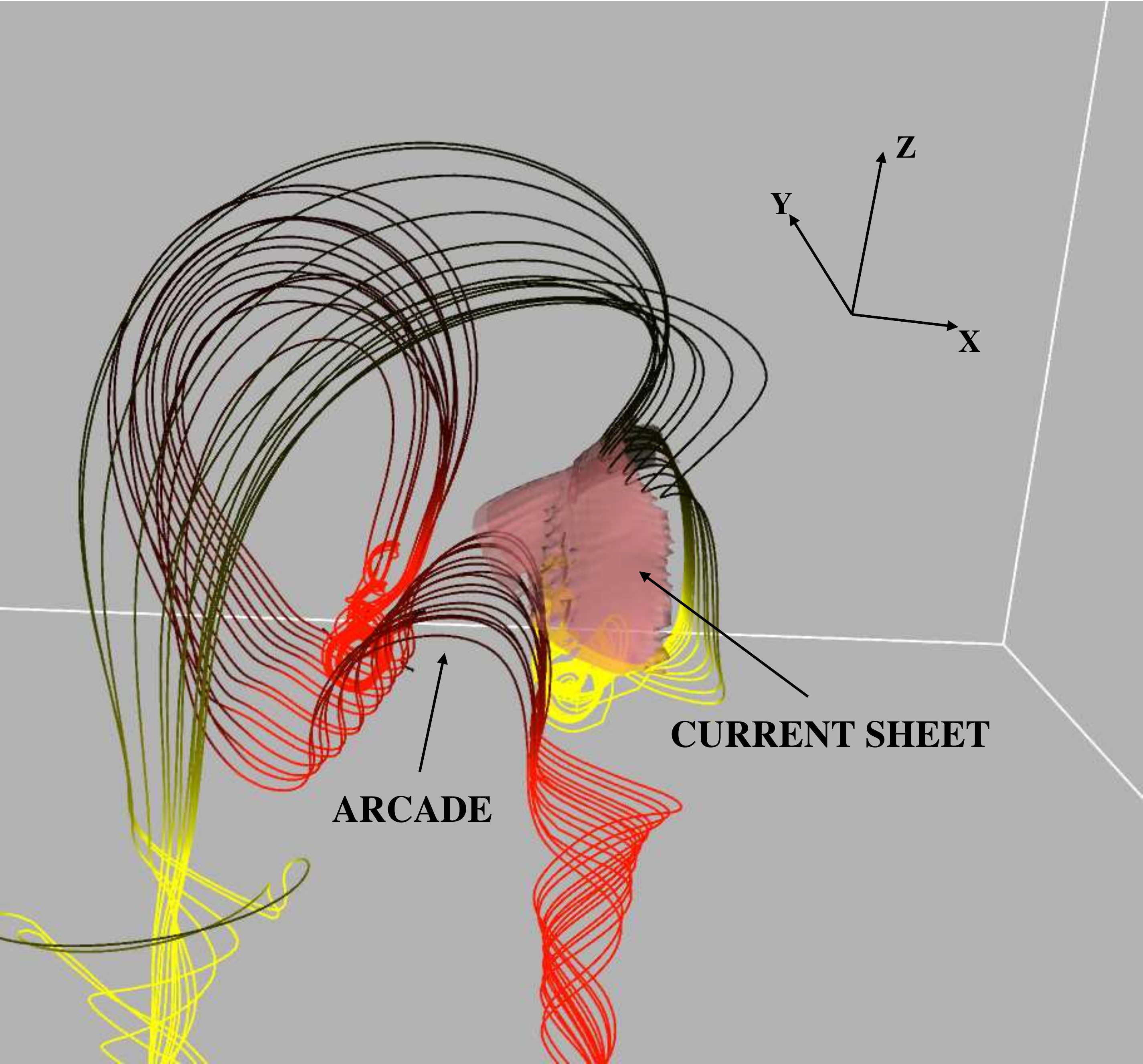}
\includegraphics[scale=0.17]{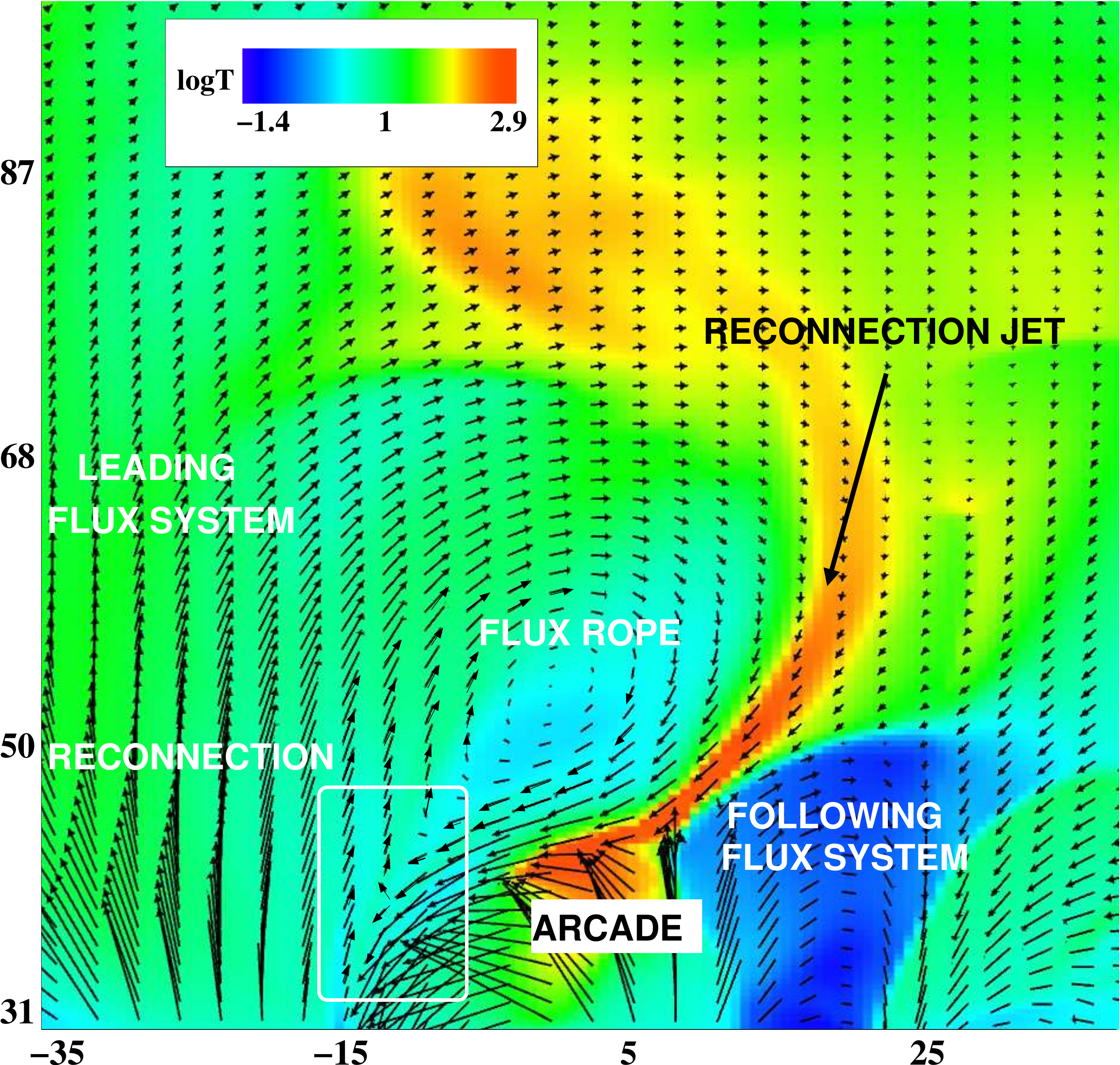}
\caption{\small {\it Left:} Visualization of the full magnetic field vector (arrows) at photospheric heights, at $t=130$. Also shown is fieldlines of the 
emerging tubes and isosurfaces of current density. The region around the asymmetry is underlined by the inset in the figure. 
{\it Middle:}Fieldline topology around the current sheet at the interface at $t=135$. The darkest the color the 
weakest the magnetic field is along the fieldlines. {\it Right:} Confined eruption 
of a flux rope at $t=140$. Superimposed is the magnetic field vector (arrows) in a vertical plane at $y=-22$. Sites of internal reconnection 
are shown in the inset rectangle of the panel.}
\label{fig1}
\end{figure*}

As the newly emerged bipolar magnetic field is moving in opposite directions along the neutral line, shearing 
of the field occurs so that the magnetic field lines lose their strong azimuthal nature 
and run nearly parallel to the neutral line. Observations have shown that most bipolar 
sheared magnetic fields in active regions can produce ejective explosions such as flares and 
CMEs \citep{Falconer01}. Also, in our simulations the twisted and sheared magnetic field adopts an overall 
\textit{sigmoidal shape} with two oppositely curved \textit{elbow-like} regions at the two (east and west) ends of the 
neutral line (as illustrated in Fig.~\ref{fig1}, left panel). The neutral line is stretched along the middle of the sigmoidal structure while the elbows are found on 
either side. During the evolution of the system, the curved region of the elbows 
consist of expanding loops of fieldlines that rise in opposite (north-east and south-west) directions. 
The parts of the elbows close to the middle of the sigmoid, shear past each other and are ready to reconnect 
if they come into contact. This magnetic field configuration with the elbow-like structures has been reported 
in observations of solar eruptive events by \citet{Moore01}.

When the second flux system emerges  at $t\approx 40$, a thin, curved, sheet-like current structure (Fig.~\ref{fig1}, middle panel) is formed just above the 
top layers of the following tube and parallel to the interface of the magnetic flux 
systems, along the y-direction. An interesting result is that already at this early stage 
of the interaction, \textit{asymmetry} 
is introduced into the system (Fig.~\ref{fig1}, left panel) by the 3D nature of the fan-like expansion of the field. 
On the west side of the emerging regions, the fieldlines of the rising systems make a contact angle 
which is larger than on the east side, leading eventually to a higher current concentration 
and more efficient reconnection at the west side of their interface.

In a similar manner to the 2.5D AHB model, soon after the build up of the 
interface current concentration \textit{plasmoid-like} structures develop inside the sheet. 
Due to the asymmetry, they are formed firstly towards the west side of the current sheet.
The cool and dense plasmoids grow and are ejected upwards and sideways, as in the AHB model, allowing 
reconnection in the diffusion regime to occur at a \textit{faster rate}. 

Two new flux systems 
are formed by the reconnected fieldlines (Fig.~\ref{fig1}, middle panel). At the \textit{upper end} of the current sheet, reconnection forms an 
overlying field that joins the two emerging systems externally, mainly from 
the north-west side of the leading system to the south-east side of the following system while it 
spreads out mostly above the west sides of the emerging systems. At the
start of the reconnection, the overlying field links the west-side flank of the leading 
tube with the east-side flank of the following tube. At the \textit{lower end} of the current sheet, reconnection 
forms a series of magnetic loops in between the two emerging systems. 
The configuration of the low-lying magnetic loops 
is an \textit{arcade-like} structure joining the leading and following tubes and is reminiscent 
of \textit{post-flare loops}. The arcade is stretched and curved along the y-direction and grows in size as 
reconnection proceeds and more fieldlines are added to the top of the arcade. The temperature 
here could be as high as $2$ MK. The arcade brightens when high velocity 
reconnection outflows, which carry hot material ejected from the current sheet, 
compress the plasma at the upper part of the arcade and increase the temperature. 

During this time, due to the emergence of the leading tube the inner fieldlines are sheared horizontally and stretched vertically. 
Hence, oppositely directed fieldlines, which are rooted to the sigmoidal neutral line, are forced into contact. This internal reconnection 
of fieldlines is similar to the `tether-cutting' mechanism \citep{Moore92}. 
In experiments with a single emerging tube, with the same initial properties, we find that internal `tether-cutting' reconnection occurs at $t\approx 180$. 
In the experiment with two emerging bipoles it occurs much  earlier at $t\approx 135$.

In the model with the two tubes, reconnection occurs first (externally) at X-type points 
at the ends of the interface current sheet.
At the lower end of the current sheet, the hot reconnected fieldlines form an arcade structure. 
The magnetic pressure inside this new arcade increases as reconnection proceeds and 
eventually exceeds the magnetic pressure 
of the leading flux system. Thus, the arcade expands sideways and pushes the south side of the leading system 
towards its neutral line. It is the combination of shearing along the neutral line, stretching due to expansion and 
the increasing magnetic pressure force of the arcade structure that leads to internal reconnection
of the fieldlines just above the 
neutral line of the leading emerging system. 

As the two emerging flux tubes continue to rise and interact, internal reconnection occurs at \textit{many positions 
along the neutral line} of the leading magnetic flux system. Due to internal reconnection, \textit{flux-rope-like structures} 
are formed inside the magnetized volume of the leading tube. Eventually, the ropes rise and erupt into the outer atmosphere. 
The eruption is due to the vigorous sideway pressure by the fast expanding arcade and the release of the magnetic tension force 
of the overlying field, after the ejection of the plasmoids out of the interface current sheet. 

A striking result is that the eruptions produced in our simulations \textit{do not appear simultaneously}. 
Also, multiple eruptions can 
be triggered from the same location inside the expanding volume of the leading system. As an example, the first formation and 
eruption of a \textit{flux rope} occurs on the south-west part of the neutral line. This part of the leading tube is closer, due to the 
asymmetry mentioned above, to the following tube. Thus, the dynamical rearrangment of the magnetic field due to reconnection 
occurs faster at this region of the leading emerging tube. 
In general, the eruptions occured in our experiment follow the direction of the overal 3D expansion of the 
emerging systems, which for the south-west side is preferentially the direction towards the following tube and for the north-east side is the 
opposite direction.
Thus, the eruption of the first flux rope is directed towards the south side (Fig.~\ref{fig1}, right panel). 
It is a confined eruption, being trapped between the envelope 
field of the leading tube and the expanding field of the upcoming tube that halts the further rise of the flux rope.
Other eruptions, which occur at 
later stages of the evolution of the emergence, are more dynamic and seem to fully escape from the corona, in a \textit{CME-like} manner. 
These eruptions 
are mainly directed towards the north side of the leading magnetic flux system.

\begin{figure}[htb]
\begin{center}
\includegraphics[scale=0.4] {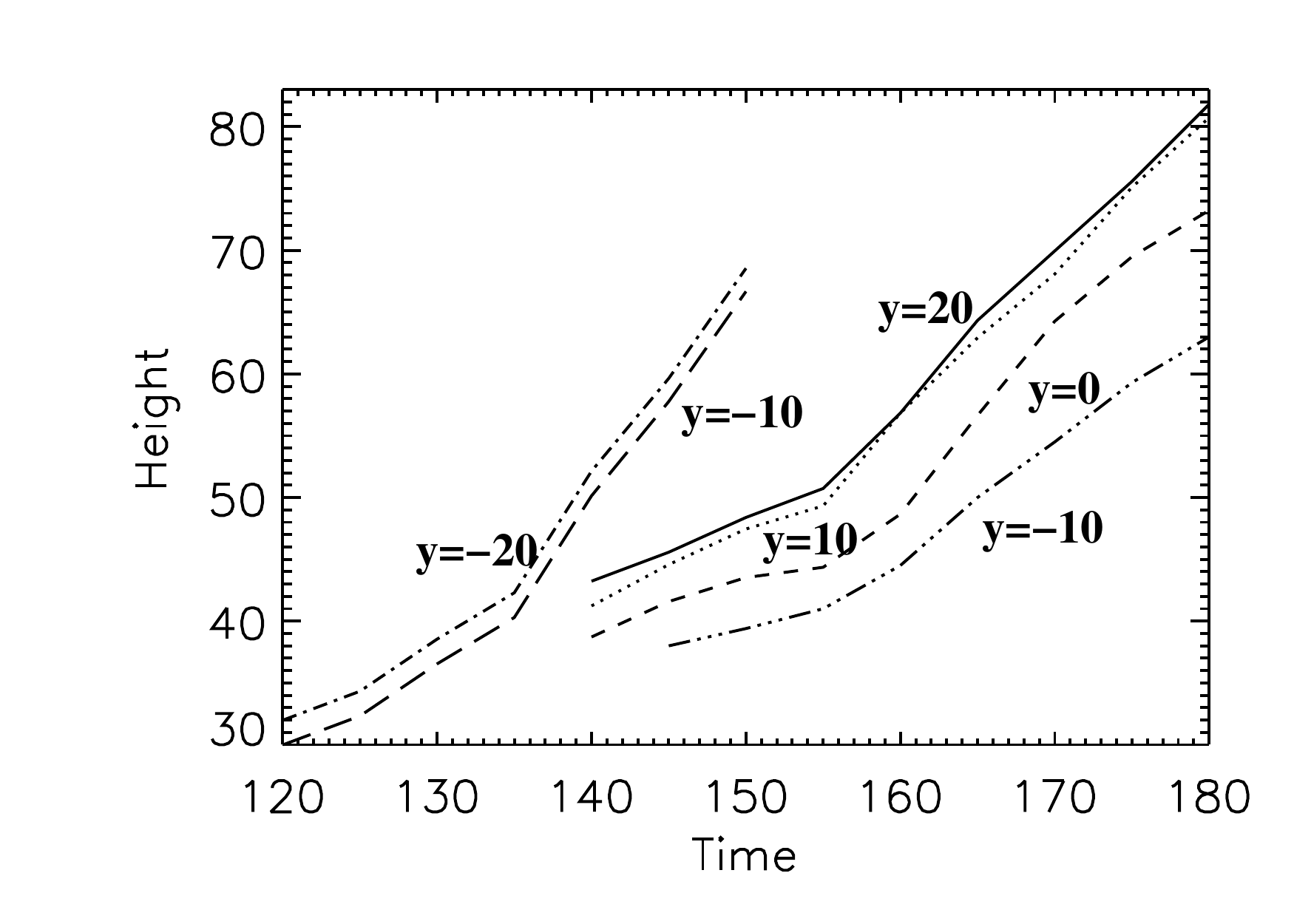}\\
\includegraphics[scale=0.4]{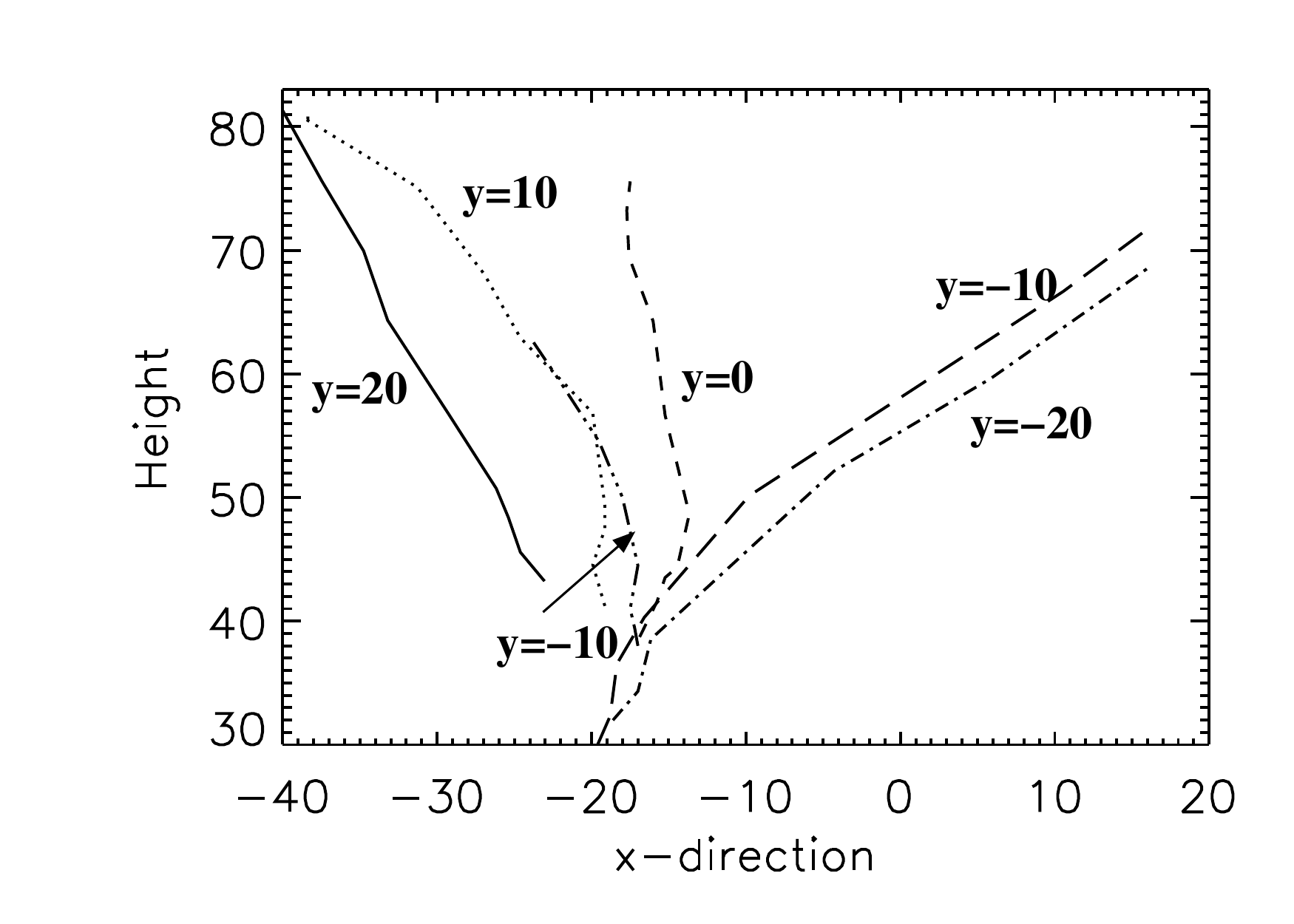}
\end{center}
\caption{\small {\it Top:} Height-time relation of the centers of the flux ropes. 
{\it Bottom:} Projection of the rising motion of the centers of the flux ropes 
on a horizontal plane.}
\label{fig2}
\end{figure}

The time evolution of the flux ropes formed \textit{inside} the leading region, can be 
followed by locating the center (or {\it O-point}) of the ropes at various  $x\, z$ cross-sections (different values of $y$) 
as functions of time. The height-time relation for 6 values of $y$ is shown in the top panel 
of Fig.~\ref{fig2}. The bottom panel is the projection of their motion onto an $x\, z$ slice. 
The confined eruption of the first flux rope is shown by the lines at $y=-20$ and $y=-10$. The center of the rope 
for $y=-20$ is lower than at $y=-10$ at all times.

There are \textit{two phases} during the eruption of the ropes. In the first,
the flux ropes are moving with moderate speeds, while in the second they are accelerated towards the outer
atmosphere. These curves are qualitatively similar to the time profiles of filament heights shown
in the observations by \citet{Sterling04, Sterling05}.
We also estimate the temperature and current density in the area below the central part of the 
erupting flux ropes where a vertical current sheet is formed due to internal reconnection.
We find that there is a slow increase in the first phase, which is followed by an intensity increase over the period of the 
fast eruption. The peak of the temperature and current density occurs when the fast rise is well underway. Since both 
quantities follow the same time evolution, it is likely that the reason for the change in the rise phase of the eruption is that 
reconnection is slower at the beginning but becomes faster, and operates with a higher growth rate, as the ropes rise.
These results are also consistent with the observations mentioned above in which soft X-ray emission is found to occur during the 
slow-rise phase of a filament's eruption and hard X-ray bursts occured after the start of the fast eruption. 

As shown above, the first formation of a flux rope and its eruption occurs in the south-west part of the leading tube.
Figure ~\ref{fig2} (bottom panel) shows that a secondary formation and eruption occurs at the same location (y=-10, dashed-triple dotted line), starting at t=145. 
Thus, \textit{at least two eruptions are triggered from the same location} within the leading emerging system until the end of the 
simulation. Further eruptions may be triggered inside the expanding magnetized volumes since the system of the 
interacting bipoles does not reach an equilibrium but evolves dynamically.
As the flux ropes erupt, they grow in radial extent as additional fieldlines reconnect at the vertical current sheet beneath them. The magnetized plasma inside the erupting ropes is a mixture of cold and hot plasma. 
The cold and dense plasma comes from the low atmosphere and is carried upwards by the tether-cutting reconnection 
forming concave upwards fieldlines. The hot plasma (up to $2$ MK) is emitted inside the envelope of the rope by the reconnection jet, 
which blasts from the 
upper end of the vertical current sheet below the rope. 
The velocity at the center of the rope is $\approx $10-15$ Km/sec$. However, the maximum upward 
velocity inside the complete volume of the rope reaches values up to $150$ Km/sec, mainly due to the 
emission of the reconnection jet. The rising motion of the flux rope can result in a CME-like eruption.

\begin{figure*}[t]
\includegraphics[scale=0.18]{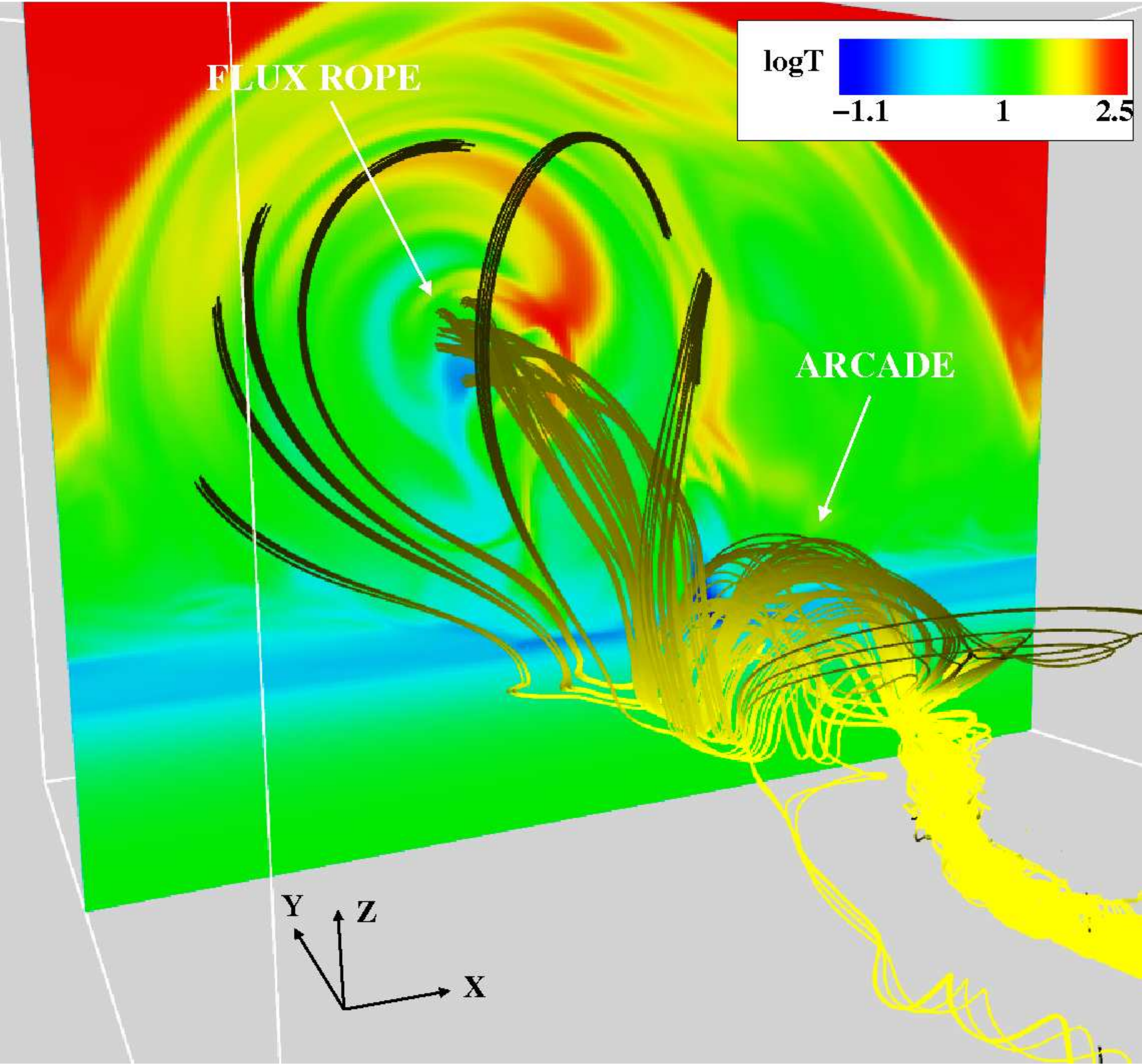}
\includegraphics[scale=0.18]{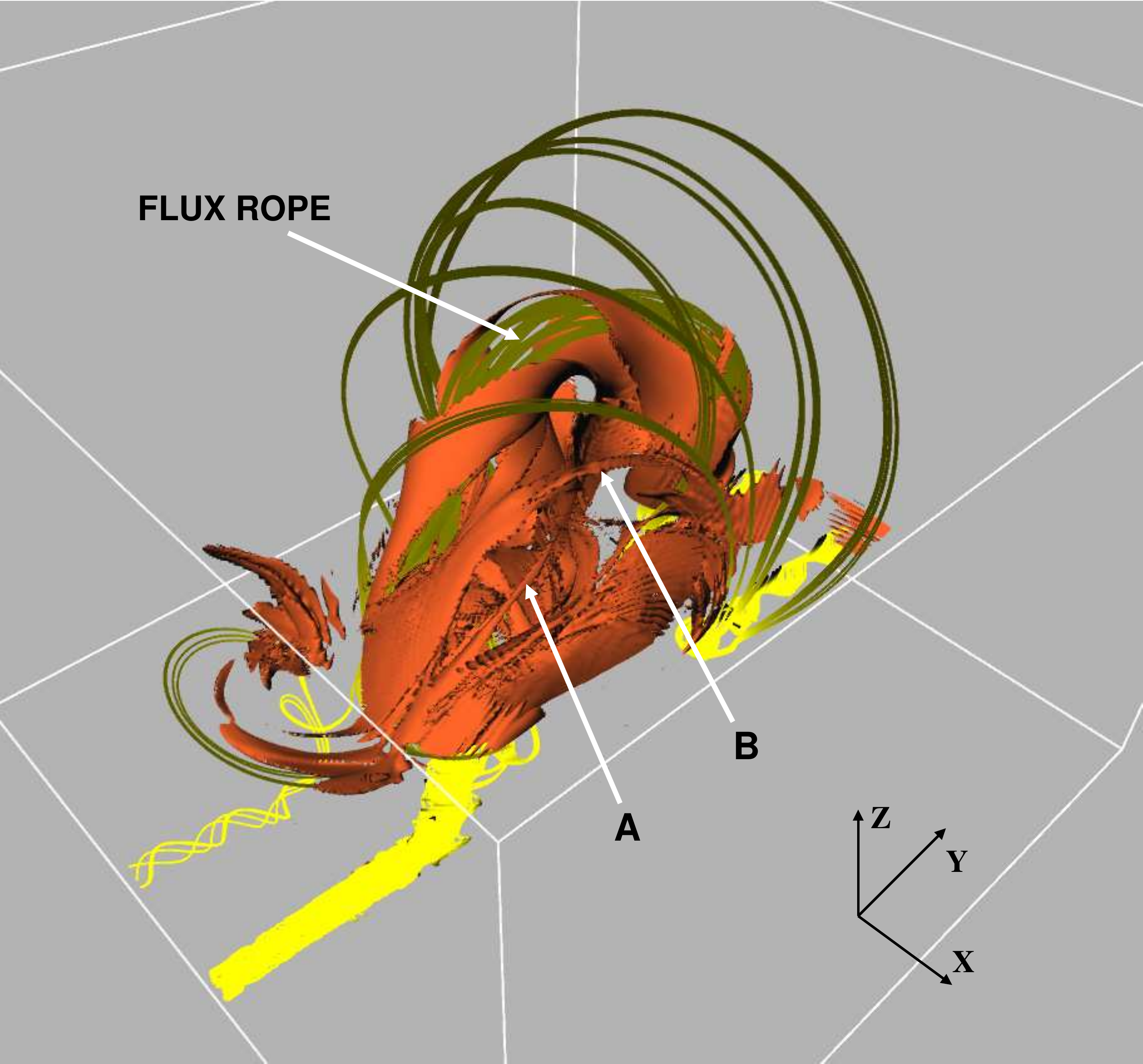}
\includegraphics[scale=0.17]{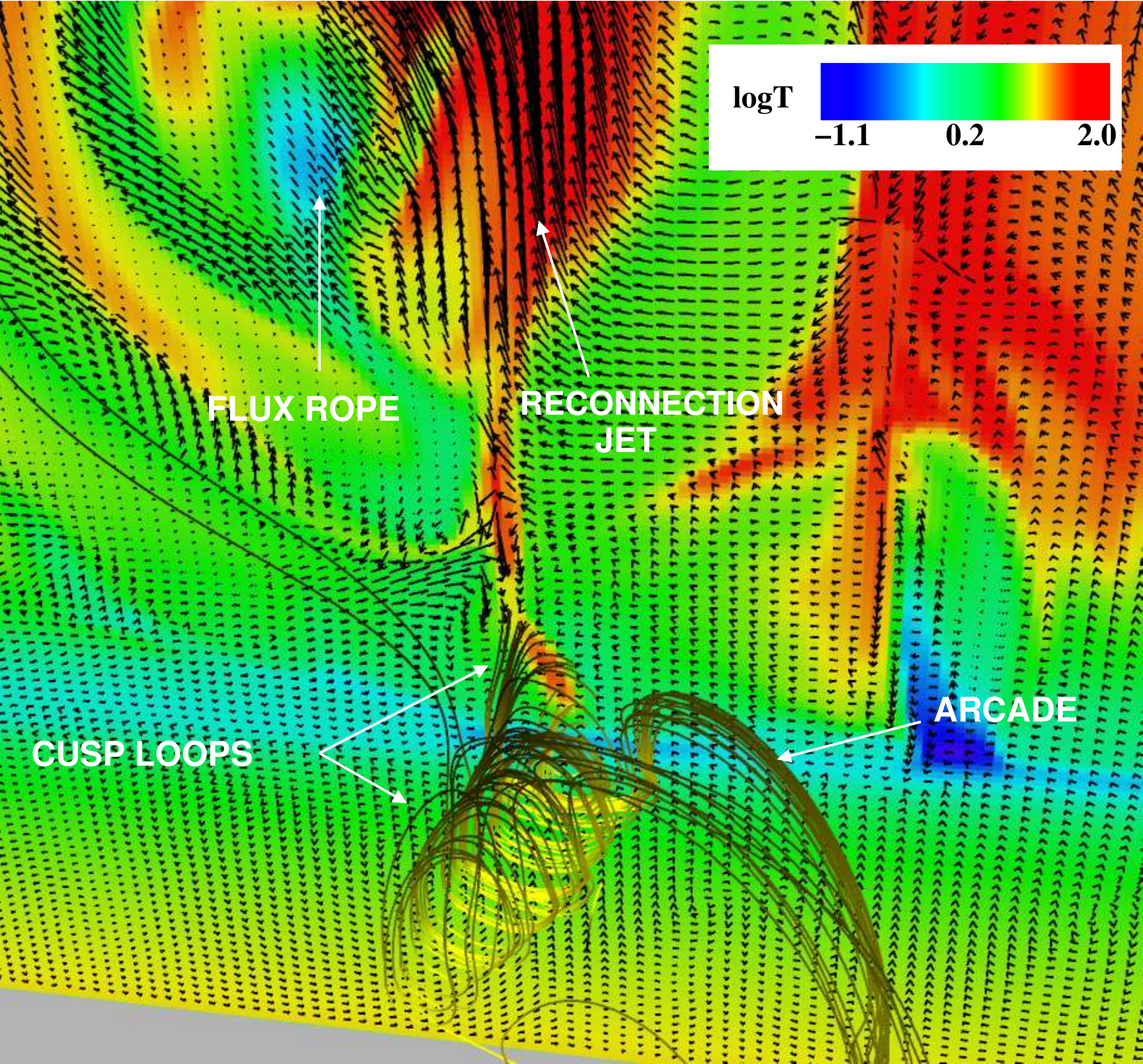}
\caption{\small {\it Left:} Ejective eruption of a flux rope at $t=175$. The temperature contours are on the plane 
$y=20$. {\it Middle:} The same as in the left panel with isosurfaces 
of current density superimposed. {\it Right:} 3D visualization of fieldlines in the region below the erupting flux rope. Superimposed on the temperature contours
is the velocity field (arrows) on $y=20$.}
\label{fig3}
\end{figure*}

Figure ~\ref{fig3} (left panel) shows the temperature distribution in a vertical plane and magnetic fieldlines, which are traced from near
the axis of the flux rope. The cold plasma is mostly concentrated near the axis of the flux rope while the hot plasma is overlying 
the axis on the right side of the eruption. The fieldlines near the axis of the rope come from \textit{both 
of the emerging flux systems}. More precisely, fieldlines from the following tube joins the leading tube, through the arcade 
structure between them before forming the erupting flux rope. These fieldlines undergo multiple reconnection events along their lengths and, thus, may end up 
in either of the flanks at the east-side boundary. 
The middle panel in Fig.~\ref{fig3} shows isosurfaces of current density, which are overplotted on the same set of fieldlines. A striking result is that long and 
twisted sheets are found close to the core of the erupting flux rope. In addition, arch-like current sheets that are less twisted and 
have different orientation to each other, are formed in the region between the emerging systems (see 
`A' and `B' in the figure). These structures, with enhanced current density, may be quasi seperatrix layers. The right panel in Figure ~\ref{fig3} is a 
blow-up of the region at the vertical current sheet below the erupting flux rope shown in the left panel. 
The fieldlines shown in the figure have 
been traced from the lower end of the vertical sheet. These are reconnected fieldlines that surround the leading flux tube's original axis. Thus, the field 
below the vertical current sheet forms a tightly wound flux tube with its axis lying parallel to the neutral line of the leading bipolar region.
The temperature rises up to $0.2$ MK at the bottom of the current sheet, as hot reconnected plasma moves downward from the sheet. These fieldlines 
are also linked: a) to fieldlines that envelope the erupting flux rope from the north side (two fieldlines on left side of figure) and b) to fieldlines that form the hot arcade 
structure between the emerging flux systems. 

Our 3D simulations show that the interaction of emerging flux tubes may lead to multiple 
eruptions of flux ropes.
A network of twisted current sheets, with an overall S-shaped configuraton, and 
hot cusp-like, arcade structures are also formed during the evolution of the system.

\acknowledgments {Computational time on the Linux clusters in
St. Andrews (STFC and SRIF funded) are gratefully acknowledged.}

\clearpage

\begin{thebibliography}{}
\bibitem[Antiochos(1998)]{Antiochos98} Antiochos, S.K. 1998,
ApJL, 502, L181
\bibitem[Arber et al.(2001)]{Arber01} Arber, T. et al. 2001,
JCoPh, 171, 151
\bibitem[Archontis, Hood \& Brady(2007)]{Archontis07} 
Archontis, V.; Hood, A. W. \& Brady, C., 2007, A\&A, 466, 367
\bibitem[Archontis et al.(2005)]{Archontis05} Archontis, V. et al. 2005,
ApJ, 635, 2
\bibitem[Chen \& Shibata(2000)]{Chen00}
Chen, P.F. \& Shibata, K., 2000, ApJ, 545, 524
\bibitem[Falconer(2001)]{Falconer01} Falconer, D.A. 2001,
JGR, 106, 25185
\bibitem[Gibson \& Fan(2006)]{Gibson+ea06}
Gibson, S. \& Fan, Y., 2006, ApJ, 637, L65
\bibitem[Jiang et al.(2007)]{Jiang07} Jiang, Y-C. et al. 2007,
CJAA, 7, 129
\bibitem[MacKay \& van Ballegooijen(2006)]{Duncan06}
Mackay, D. H. \& van Ballegooijen, A. A., 2006, ApJ, 641, 577
\bibitem[Manchester et al.(2004)]{Manchester+ea04} Manchester, W.IV. et al. 2004,
ApJ, 610, 588
\bibitem[Moore et al.(2001)]{Moore01} Moore, R.L. et al. 2001,
ApJ, 552, 833
\bibitem[Moore \& Roumeliotis(1992)]{Moore92}
Moore, R.L. \& Roumeliotis, G., 1992, Eruptive Solar Flares. 
Proceedings of IAU Colloquium 133.
\bibitem[Murray et al.(2006)]{Murray06} Murray, M.J. et al. 2006,
A\&A, 460, 909
\bibitem[Sturrock (1989)]{Sturrock89} Sturrock, P.A. 1989,
Solar Physics, 121, 387
\bibitem[Sterling \& Moore(2004)]{Sterling04}
Sterling, A.C. \& Moore, R.L., 2004, ApJ, 613, 1221
\bibitem[Sterling \& Moore(2005)]{Sterling05}
Sterling, A.C. \& Moore, R.L., 2005, ApJ, 630, 1148
\bibitem[T\"{o}r\"{o}k \& Kliem(2005)]{Torok05}
T\"{o}r\"{o}k, T. \& Kliem, B., 2005, ApJ, 630, L97
\bibitem[van Ballegooijen \& MacKay(2007)]{Adrian07}
van Ballegooijen, A. A. \& Mackay, D. H., 2007, ApJ, 659, 1713
\end{thebibliography}
\end{document}